\documentclass[a4paper,12pt]{article}
\usepackage{epsfig}
\makeatletter
\def\vereq#1#2{
\lower3pt\vbox{\baselineskip1.5pt \lineskip1.5pt
\ialign{$\m@th#1\hfill##\hfil$\crcr#2\crcr\sim\crcr}}}
\makeatother

\begin{document}

\begin{titlepage}
\begin{center}
\hfill    CERN-PH-TH/2004-038\\
~{} \hfill hep-ph/0402yyy\\
\vskip 1cm

{\large \bf  Non-Factorizable Phases, Yukawa Textures \\ and 
the Size of \boldmath $\sin 2 \beta $}

\vskip 1cm

G. C. Branco\footnote{On leave of absence from 
Departamento de F\'\i sica,
Instituto Superior T\' ecnico, Av. Rovisco Pais, P-1049-001, Lisboa,
Portugal.}\footnote{E-mail: gustavo.branco@cern.ch and
gbranco@alfa.ist.utl.pt},
M. N. Rebelo$^ \ast$\footnote{E-mail: margarida.rebelo@cern.ch and
rebelo@alfa.ist.utl.pt} 
and  
J. I. Silva-Marcos$^ \ast$\footnote{E-mail: Joaquim.Silva-Marcos@cern.ch}

\vskip 0.05in

{\em Department of Physics,
Theory Division, CERN, \\ CH-1211 Geneva 23, Switzerland.}
\end{center}

\vskip 3cm

\begin{abstract}
We emphasize the crucial r\^ ole played by non-factorizable
phases in the analysis of the Yukawa flavour structure performed
in weak bases with Hermitian mass matrices and with vanishing $(1,1)$ 
entries.
We show that non-factorizable phases are important  in order to
generate a sufficiently large $\sin 2 \beta $. A method is suggested to 
reconstruct the flavour structure of Yukawa couplings from
input experimental data both in this  
Hermitian basis and in a non-Hermitian basis with a
maximal number of texture zeros. The corresponding Froggatt--Nielsen
patterns are presented in both cases.
\end{abstract}

\end{titlepage}

\newpage 

\section{Introduction}

The pattern of quark mass matrices and the implied flavour 
structure of Yukawa couplings may play a crucial r\^ ole in providing
evidence for the possible existence of an underlying family symmetry.
The search for a family symmetry has become more urgent in view of the 
significant improvement of our knowledge \cite{dois} of the 
Cabibbo--Kobayashi--Maskawa ($V_{CKM}$) matrix 
and the value of quark masses. Recent developments in the theory of
heavy quarks, as well as in experimental techniques, have led
to a more precise knowledge of the elements of $V_{CKM}$, 
with $|V_{us}|$ and  $|V_{cb}|$ known with errors at the 
2$\% $ level \cite{dois}; the uncertainty on $|V_{ub}|$
is of about  13$\% $, but there is hope to achieve
a precision below the 10$\% $ level in the near future 
\cite{Battaglia:2004ti}. On the other hand, the measurements of
$\sin (2 \beta ) $ by B-factories at BaBar \cite{Eigen:2003rk} 
and Belle \cite{Abe:2003yu}
have reached the  6.5$\% $ level \cite{updated}, with more 
precise measurements expected in the near future. All these
developments, combined with an improvement in our knowledge of quark 
masses arising from lattice QCD \cite{Gupta:2001cu}, constitute
a great challenge for any theory of flavour.

The major difficulty
one encounters when attempting to reconstruct the flavour structure of 
Yukawa couplings from experimental data stems from the fact that even a 
precise knowledge of the value of the six quark masses and the four 
physical parameters of $V_{CKM}$ does not lead to a unique reconstruction
of the Yukawa couplings. This reflects the freedom one has in the Standard 
Model (SM) of making weak basis (WB) tranformations, which leave the 
charged currents diagonal and real, but change the flavour structure 
of Yukawa couplings. For a given set of precise 
experimental input data, there is 
an infinite set of Yukawa structures compatible with the data and
related to each other through WB   
transformations. Even if it is assumed that there is indeed 
a family symmetry
chosen by nature, the question remains of discovering  in which WB  
that symmetry will be ``transparent''. 

In the literature, there are various approaches 
to the Yukawa puzzle \cite{rev} including,
for example, the assumption of a set of texture zeros 
\cite{Ramond:1993kv}--\cite{Kim:2004ki} or the hypothesis of 
universality of the strength of Yukawa couplings 
(USY) \cite{usy}. In all these 
approaches there is the implicit adoption of a specific WB, where the proposed 
feature of Yukawa couplings is manifest. 
  
In this paper, we discuss possible ways of determining the flavour structure
of Yukawa couplings from input experimental data, in spite of the 
difficulty described above. First we choose a 
WB where the quark mass matrices are Hermitian
and hierarchical, with a vanishing $(1,1)$ element both in the up 
and down quark mass matrices. It is worth emphasizing that this
is a choice of WB, not implying any loss of generality. Indeed it has 
been shown \cite{Branco:1999nb} that
starting from arbitrary complex matrices $M_u$, $M_d$ one can 
always make WB transformations
that render both $M_u$, $M_d$ Hermitian and 
with vanishing $(1,1)$ elements.
There are various motivations for considering this WB :

i) We will argue that although,  this choice of WB 
by itself, has of course, no 
physical implications,
it leads,  when combined with a requirement of naturalness,
to an understanding of the size of the Cabibbo mixing.
This ``naturalness'' condition consists of considering 
that the smallness of $|V_{ub}|$ results from the smallness 
of $|U^u_{31}|$, $|U^d_{13}|$ and not from the cancellation of 
the contributions to  $V_{ub}$ arising from  
$U^u$ and $U^d$, the two unitary matrices
that diagonalize  $M_u$, $M_d$.

ii) Most, if not all, of the ans\" atze considered in the literature,
based on Hermitian and hierarchical quark mass matrices, have the
above two texture zeros. \\

In our analysis, we classify the phases appearing in $U^u$ and $U^d$
into two categories, namely factorizable and non-factorizable
phases. We show that, in the framework of our chosen WB, the existence
of at least one non-factorizable phase is essential in obtaining
a sufficiently large value of $\sin (2 \beta ) $, to be consistent 
with experiment. This result is specially important in view of the fact
that in most of the ans\" atze considered in the literature, based
on Hermitian and hierarchical mass matrices, the non-factorizable
phases can be rotated away.

The number of free parameters
in this WB exceeds the number of measurable quantities and therefore
we cannot reconstruct $M_u$, $M_d$ just from input data. In order 
to achieve this task, we propose a set of well motivated
naturalness criteria,  which render the reconstruction possible. 
The resulting quark mass patterns then have only two texture
zeros and it is possible to write the remaining small entries
in terms of powers of a small parameter.
The fact that we have only two texture zeros should not be interpreted 
as disfavouring the presence of a family symmetry.
Indeed, the modern approach to texture zeros allows for a small deviation
of the exact relations resulting from strict zero entries, 
since it is to be expected that family symmetries only require
texture zeros to be approximate. Furthermore, renormalization group
running, although usually leading to small effects, would not
keep exact zeros at all scales. A possible mechanism producing 
structures with small but non-vanishing entries is the spontaneous 
breaking of a family symmetry through new scalar fields at high
energies, leading to effective Yukawa couplings at low energies via the
Froggatt--Nielsen mechanism \cite{Froggatt:1978nt}. 
The point of our analysis is deriving this Froggatt--Nielsen 
pattern from input data.

Most of the analysis presented in this article is made in the
framework of the above described Hermitian WB. For completeness,
we will also consider another possible approach for reconstructing
$M_u$, $M_d$ from input experimental data. This proposal 
arises as an attempt at answering the following question: Starting from
arbitrary quark mass matrices  $M_u$, $M_d$, what is the maximal 
number of zeros that can be achieved, by making WB transformations?
We will show that if we do not require $M_u$, $M_d$ to be Hermitian, 
the maximal number of zeros is nine. In this WB, $M_u$, $M_d$ contain 
a total of ten free parameters, nine real numbers and one physical
phase. Since the number of free parameters equals the number of 
measurable quantities (i.e., the six quark masses and the four 
physical parameters contained in $V_{CKM}$), it is possible to fully
reconstruct $M_u$, $M_d$ from experiment. This is an interesting 
feature of this WB. Its possible drawback is the fact that $M_u$, $M_d$
are not Hermitian and furthermore 
that they are not treated on an equal footing.
An interesting consequence of our analysis is that, in this WB, both
quak mass matrices are of the nearest-neighbour-interaction (NNI) form, 
but the experimental data require a strong deviation from
hermiticity. This result was to be expected, since the NNI basis
\cite{Branco:1988iq} with hermiticity leads to the Fritzsch ansatz 
\cite{Fritzsch}, which is known to
be ruled out by experiment.

\section{Factorizable and non-factorizable phases}

\setcounter{equation}{0}

\subsection{Definitions}

It is well known that one can choose, without loss of generality, a
weak basis (WB) where both the up and down quark mass matrices $M_{u,d}$ are
Hermitian and therefore can be parametrized in the form: 
\begin{equation}
\label{uud} M_u = U^u \ D_u \ U^{u \dagger }  \ ;\ \ \ 
M_d = U^d \ D_d \ U^{d \dagger } \ ,
\end{equation}
where $D_{u,d}$ denote real diagonal matrices and the 
unitary matrices $U^{u,d}$
can be written as: 
\begin{equation}
\label{puk}U^u ={P_u}\ U{_o^u}\ K_u\ ;\ \ \ U^d ={P_d}\ U{_o^d}\ K_d
\end{equation}
with $P_u\equiv {\rm diag}[e^{i\phi _1^u},1,e^{i\phi _3^u}]$ , 
$K_u\equiv {\rm diag}[e^{i\alpha _1^u},e^{i\alpha _2^u},e^{i\alpha _3^u}]$,
with analogous expressions for $P_d$, $K_d$. The matrices $U{_o^{u,d}}$ are
unitary with only one phase each and can be parametrized using, for example,
the standard parametrization \cite{Hagiwara:fs}. These phases, which we
denote by $\sigma _u$, $\sigma _d$ are {\bf non-factorizable phases}, in the
sense that they cannot be removed by redefinitions of $P_{u,d}$, $K_{u,d}$.
The CKM matrix is then given by: 
\begin{equation}
\label{ckm}V_{CKM}=U{_o^u}\ ^{\dagger }P\ U{_o^d},
\end{equation}
with $P\equiv {\rm diag}[e^{i{\phi _1}},1,e^{i{\phi _3}}]$ where $\phi
_1=\phi _1^d-\phi _1^u$, $\phi _3=\phi _3^d-\phi _3^u$. We have omitted in 
$V_{CKM}$ the matrices $K_u$, $K_d$ since their phases can be eliminated
through redefinitions of up and down quark fields. We call the phases 
$\phi _1$ and $\phi _3$ {\bf factorizable phases} in contrast with the {\bf 
non-factorizable phases} $\sigma _u$, $\sigma _d$ 
contained in $U{_o^u}$, $U{_o^d}$. 
Needless to say, neither $U{_o^u}$ nor $U{_o^d}$ are measurable
quantities, since only $V_{CKM}$ appears in the charged weak currents. Of
course, for three fermion generations $V_{CKM}$ will contain only one
physical phase, which is a complicated function of the factorizable and
non-factorizable phases. CP violation through the KM mechanism can be
generated even in the limit where only the factorizable phases $\phi _1$, 
$\phi _3$ are present, as well as in the limit where 
only the non-factorizable
phases are non-vanishing. 

Although the unitary matrices 
$U{_o^u}$, $U{_o^d}$
are not measurable, they are useful in the discussion of specific flavour
structures for $M_u$, $M_d$. 
In the following, we will show that in most of the
flavour structures considered in the literature, 
the structure of $M_u$, $M_d$ are such that there are 
no non-factorizable phases. This means that for
these textures, the unitary triangles corresponding to $U{_o^u}$, $U{_o^d}$
collapse to a line. For these flavour structures, CP violation arises
exclusively from the factorizable phases $\phi _1$ and $\phi _3$. We will
see that this feature plays a crucial r\^ole in preventing these flavour
textures from generating a sufficiently large value for $\sin (2\beta )$,
recently measured with significant accuracy by BaBar and Belle.

\subsection{A class of ans\" atze without non-factorizable phases}

Next, we will show that there is a large class 
of ans\"atze for $M_u$, $M_d$, where there are no non-factorizable 
phases. This includes all  five
texture zeros classified by Ramond, Roberts and Ross \cite{Ramond:1993kv},
as well as the four texture zeros considered in \cite{Branco:1999tw}, \cite
{Fritzsch:2002ga}. In order to see how this result comes about, note that
from Eq.~(\ref{uud}) one readily obtains: 
\begin{equation}
\label{imm}\left| {\rm Im}[(M_u)_{12}(M_u)_{23}(M_u)_{13}^{*}]\right|
=(m_3^u-m_2^u)(m_2^u-m_1^u)(m_3^u-m_1^u)~|{\rm Im}(Q_u)|, 
\end{equation}
where $Q_u$ denotes any rephasing-invariant quartet constructed with the
elements of $U^u$. Obviously, an analogous result holds 
for $M_d$. From Eq.~(\ref{imm}), it follows that if at least one of 
the off-diagonal elements of $M_{u,d}$ vanishes then 
${\rm Im}(Q_{u,d})=0$, which in turn implies the
absence of non-factorizable phases in $U^{u,d}$. It is seen from Table 1 of
Ref. \cite{Ramond:1993kv} that all  five texture zeros classified there
and the four texture zeros analysed in \cite{Branco:1999tw}, \cite
{Fritzsch:2002ga} have at least one vanishing off-diagonal element and
that therefore there are no non-factorizable phases in these textures.

\section{The search for an appropriate \\ Hermitian weak basis}

In this section, we address the question of discovering an appropriate
Hermitian WB for reconstructing the quark mass matrices $M_u$, $M_d$ from
input experimental data. The hope is to get from these reconstructed
matrices a hint of the underlying symmetry. Of course, we have to define
what is meant by an ``appropriate'' WB. A criterion for the choice of WB
could be, for example, the requirement of having a number of free parameters
equal to the number of measurable quantities. This would mean a total of ten
free parameters, to be determined from the knowledge of the six quark masses
and the four physical parameters contained in $V_{CKM}$. Actually, there are
two well known Hermitian WBs 
that satisfy the above criterion, namely the
WB where $M_u$ is diagonal real and $M_d$ Hermitian, 
and the one where it is $M_d$ that is diagonal real and 
$M_u$ Hermitian. It is clear that each one of
these two WBs has ten free parameters. For example, in the WB where $M_u$ is
diagonal real, one would have as parameters the three up-quarks, the six
real parameters contained in the Hermitian 
$M_d$, and the physical phase 
$\arg [(M_d)_{12}(M_d)_{23}(M_d)_{13}^{*}]$, 
which is the only phase that
cannot be rephased away. The reconstruction of $M_d$ from experimental data
is then straightforward, since one has $M_d=V_{CKM}\cdot {\rm diag}
[m_d,m_s,m_b]\cdot V_{CKM}^{\dagger }$. The obvious disadvantage of these
two WBs is the fact that in each one of them, $M_u$, $M_d$ are not treated
on an equal footing. 

We will argue that 
the WB where $M_u$, $M_d$ are Hermitian
and furthermore $(M_u)_{11}=(M_d)_{11}=0$ is an appropriate basis for
reconstructing the quark mass matrices from input data. 
This WB treats $M_u$, $M_d$ on an equal footing, 
but it has, of course, the disadvantage of having
more free parameters than measurable quantities, thus rendering the
reconstruction somewhat ambiguous. We argue below that such a
reconstruction is possible, provided some naturalness
criteria are introduced.

\subsection{The {\protect\boldmath $(M_u)_{11} = (M_d)_{11} = 0$} weak basis}

Starting with arbitrary quark mass matrices, it has been shown that one can
always make WB transformations such
that $(M_u)_{11}=(M_d)_{11}=0$. 
From Eqs.~(\ref{uud}), we obtain in this basis: 
\begin{equation}
\label{dsb}
\begin{array}{c}
m_u|U_{11}^u|^2-m_c|U_{12}^u|^2+m_t|U_{13}^u|^2=0 \\  
\\ 
m_d|U_{11}^d|^2-m_s|U_{12}^d|^2+m_b|U_{13}^d|^2=0 \ ,
\end{array}
\end{equation}
where we have made the identification $m_1^u\equiv m_u$, $m_2^u\equiv -m_c$, 
$m_3^u\equiv m_t$ and analogously for the down quark sector. 
Since the $(1,1)$
zeros only reflect a choice of basis, they do not have, by themselves, any
physical implications. However, when combined with a reasonable assumption
on the smallness of $|U_{13}^u|$, $|U_{13}^d|$ one is led to the prediction: 
\begin{equation}
\label{sqr}\frac{|U_{12}^u|}{|U_{11}^u|}\approx \sqrt{\frac{m_u}{m_c}}\ ;\ \
\ \frac{|U_{12}^d|}{|U_{11}^d|}\approx \sqrt{\frac{m_d}{m_s}}.
\end{equation}
In order to qualify what we mean above by ``reasonable assumption'', let us
recall that: 
\begin{equation}
\label{vub}V_{ub}=U{_{11}^{u*}}U_{13}^d+U{_{21}^{u*}}
U_{23}^d+U_{31}^{u*}U_{33}^d \ .
\end{equation}
The experimental value of $|V_{ub}|$ tells us that $|V_{ub}|$ is of order $%
m_d/m_b$ and therefore from Eq.~(\ref{vub}), assuming $|U_{11}^u|=O(1)$ and
barring unnatural cancellations in the three terms contributing to $|V_{ub}|$%
, one is led to the conclusion that $|U_{13}^d|$ cannot be much larger than $%
m_d/m_b$ . Assuming that $M_u$, $M_d$ have analogous flavour structures, one
is led to the following order of magnitude for $|U_{13}^d|$, $|U_{13}^u|$: 
\begin{equation}
\label{u13}|U_{13}^d|=O(m_d/m_b)\ ;\ \ \ |U_{13}^u|=O(m_u/m_t)
\end{equation}
From Eq.~(\ref{u13}) it then follows that $m_t\ |U_{13}^u|^2\approx
m_u\left( \frac{m_u}{m_t}\right) $ , $m_b \ |U_{13}^d|^2\approx m_d\left( 
\frac{m_d}{m_b}\right) $, thus implying that the third terms in Eqs. (\ref
{dsb}) are entirely negligible, which in turn leads to Eq.~(\ref{sqr}). From
this equation, one then obtains: 
\begin{equation}
\label{vus}|V_{us}|=\left| \sqrt{\frac{m_d}{m_s}}+e^{i\alpha }\sqrt{\frac{m_u%
}{m_c}}\right| \ ,
\end{equation}
with $\alpha \equiv \arg (U_{11}^uU_{22}^dU_{21}^{u*}U_{12}^{d*})$. It is
well known that Eq.~(\ref{vus}) can be obtained \cite{rev}
in the framework of
models with a specific set of texture zeros for $M_u$, $M_d$. The imposition
of these sets of zeros goes beyond a choice of WB. The point of the above
discussion is that one may arrive at Eq.~(\ref{vus}) with much weaker
assumptions, namely the choice of the $(1,1)=0$ WB (we emphasize that this
is always possible to achieve starting with arbitrary $M_u$, $M_d$), together
with some reasonable qualitative assumptions on the size of $|U_{13}^u|$, $%
|U_{13}^d|$. The fact that Eq.~(\ref{vus}) is in good agreement with
experiment can be interpreted as an indication that the  $(1,1)=0$ WB is
a good basis to derive the structure of $M_u$, $M_d$ and thus that of Yukawa
couplings, from experimental input. This question will be addressed in
Section 4.

\subsection{The experimental value of 
{\protect\boldmath $\sin (2 \beta )$}
and  \\ the need for non-factorizable phases}

We point out that in schemes where the $(1,1)=0$ basis is used and if one
adopts some naturalness requirements, at least one non-factorizable phase is
needed in order to achieve a sufficiently large value of $\sin (2\beta )$.
In order to show how this result is obtained, let us first consider the case
where $U_o^u$, $U_o^d$ in Eq.~(\ref{ckm})do not contain any non-factorizable
phases. This is equivalent to assuming that: 
\begin{equation}
\label{imq}{\rm Im}(Q_u)={\rm Im}(Q_d)=0 \ ,
\end{equation}
where $Q_u$, $Q_d$ are any of the rephasing invariant quartets constructed
from $U_o^u$, $U_o^d$. As we have previously emphasized, the conditions of
Eq.~(\ref{imq}) are satisfied in any scheme where $M_u$, $M_d$ are Hermitian
and there is at least one off-diagonal texture zero in both $M_u$ and $M_d$.
In this case, the matrices $U_o^u$, $U_o^d$ 
are orthogonal real matrices,
which we denote by $O^u$, $O^d$. The CKM matrix can then be written in terms
of $O^u$, $O^d$ and the factorizable phases $\phi _1$, $\phi _3$ leading,
for example, to: 
\begin{eqnarray}
V_{cd} &  = & e^{i \phi_{1} } O^u_{12}O^d_{11} + O^u_{22}O^d_{21} +
e^{i \phi_{3}} O^u_{32}O^d_{31} \nonumber \\
V_{cb} & = & e^{i \phi_{1} } O^u_{12}O^d_{13} + O^u_{22}O^d_{23} +
e^{i \phi_{3}} O^u_{32}O^d_{33} \nonumber \\
V_{td} & = & e^{i \phi_{1} } O^u_{13}O^d_{11} + O^u_{23}O^d_{21} +
e^{i \phi_{3}} O^u_{33}O^d_{31} \label{lot} \\
V_{tb} & = &  e^{i \phi_{1} } O^u_{13}O^d_{13} + O^u_{23}O^d_{23} +
e^{i \phi_{3}} O^u_{33}O^d_{33} \ . \nonumber
\end{eqnarray}
In order to estimate the size of 
\begin{equation}
\label{beta}\beta \equiv \arg (-V_{cd}V_{tb}V_{cb}^{*}V_{td}^{*})
\end{equation}
we have to make a set of assumptions:

(i) From experiment, we know that $V_{CKM}\equiv U^{u \dagger }\cdot
U^d\approx {1\>\!\!\!{\rm I}}$ . We assume that this results from having
both $U^u\approx {1\>\!\!\!{\rm I}}$, $U^d\approx {1\>\!\!\!{\rm I}}$ rather
than from unnatural cancellations in the contributions of $U^u$, $U^d$ to $%
V_{CKM}$.

(ii) We assume that the magnitude of the off-diagonal matrix elements 
of $U^u$ are significantly smaller than the corresponding 
ones in $U^d$, i.e. that $|U_{ij}^u|\ll |U_{ij}^d|$ 
for $i\neq j$. This is certainly true for all
flavour models with analogous texture zeros for $M_u$, $M_d$, where the
mixing angles are expressable in terms of quark mass ratios. Since the
hierarchy of quark masses are stronger in the up-quark sector (e.g. $%
m_u/m_t\ll m_d/m_b$, $m_u/m_c\ll m_d/m_s$) this leads to smaller mixing
angles in that same sector. 

\noindent
Taking into account the above reasonable assumptions, we obtain to an
excellent approximation: 
\begin{equation}
\label{meq}
\begin{array}{lll}
\arg V_{tb}\simeq \phi _3 &  & \arg V_{td}^{*}\simeq -\phi _3+\epsilon _{td}
\\ 
\arg V_{cb}^{*}\simeq \epsilon _{cb} &  & \arg (-V_{cd})\simeq \arctan
( \frac{O_{12}^u\sin \phi _1}{O_{21}^d} )\ ,
\end{array}
\end{equation}
where 
\begin{eqnarray}
\epsilon _{td} = \arctan \left[ \frac{O^u_{23}O^d_{21} \sin \phi_{3}}
{O^u_{33}O^d_{31} + O^u_{23}O^d_{21} \cos \phi_{3} } \right] \\
\epsilon _{cb} = \arctan \left[ \frac{-O^u_{32}O^d_{33} \sin \phi_{3}}
{O^u_{22}O^d_{23} + O^u_{32}O^d_{33} \cos \phi_{3} } \right] \ . 
\end{eqnarray}
From Eq.~(\ref{meq}), it follows that : 
\begin{equation}
\label{e13}\beta \simeq \arctan \left( \frac{O_{12}^u\sin (\phi _1)}
{O_{21}^d} \right)+\epsilon _{td}+\epsilon _{cb} \ .
\end{equation}
In the framework of our assumptions (i), (ii), 
it can readily be seen that $\epsilon _{td}$, 
$\epsilon _{cb}$ are small. For example, one 
expects $O_{31}^d$ to give the dominant contribution 
to $V_{td}$ and therefore to be
of order $\lambda ^3$ ($\lambda $ denoting as usual the Cabibbo angle). On
the other hand, one expects $O_{21}^d\simeq \lambda $, but $O_{23}^u$
significantly smaller than $\lambda ^2$, because of the strong up quark
hierarchy. As a result one has $\epsilon _{td}\ll 1$. Similar arguments
apply to $\epsilon _{cb}$ so that $\epsilon _{cb}\ll 1$. Therefore, the
dominant contribution to $\beta $ is given by the first term in Eq.~(\ref
{e13}), which arises from $\arg (-V_{cd})$. In order to evaluate $\arg
(-V_{cd})$, one has to know the value of $\phi _1$. This is essentially
fixed by the magnitude of $V_{us}$, which is given by : 
\begin{equation}
\label{aus}|V_{us}|=\left| e^{i\phi
_1}O_{11}^uO_{12}^d+O_{21}^uO_{22}^d+e^{i\phi _3}O_{31}^uO_{32}^d\right|
\simeq \left| e^{i\phi _1}O_{11}^uO_{12}^d+O_{21}^uO_{22}^d\right| \ . 
\end{equation}
Using Eq.~(\ref{sqr}), together with the central values for the quark mass
ratios and the experimental value of $|V_{us}|$: 
\begin{equation}
\label{exp}\frac{m_d}{m_s}=\frac 1{20}\ ,\ \ \ \frac{m_u}{m_c}=\frac 1{325}\
,\ \ \ |V_{us}|=0.2196\pm 0.0026 \ ,
\end{equation}
one obtains: 
\begin{equation}
\label{cir}\phi _1= - 1.77\ {\rm {rad}\ \ (- 101^{\circ }).}
\end{equation}
Note that $|V_{us}|$ only fixes $|\phi_1|$. We have opted for
a negative value for  $\phi_1$, so that the leading contribution
to $\beta $ in Eq.~(\ref{e13}) be positive (recall that $O^d_{21}$
is negative in the standard parametrization).
From Eqs.~(\ref{e13}) and
(\ref{cir}) one obtains for the leading contribution
to $\beta $: 
\begin{equation}
\label{lea}\beta _{{\rm {leading}}}\approx 0.24\ {\rm {rad}\ \
(14^{\circ }).}
\end{equation}
Note that at this stage we have neglected the contributions to $\beta $
arising from $\epsilon _{td}$, $\epsilon _{cb}$. 
\begin{figure}[t]
\vspace{2.0truecm} \centerline{\ \epsfysize=8.0truecm \epsffile{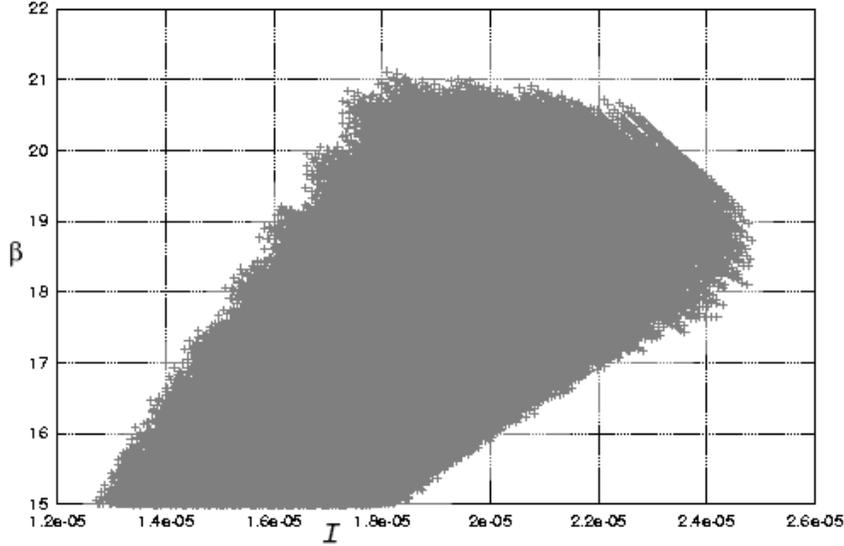}}
\caption{$\beta $ as a function of $I$, without non-factorizable phases
and quark masses varying within the allowed range}
\label{bej2}
\end{figure}
In order to evaluate the size of these contributions we write: 
\begin{equation}
\label{que}O_{13}^d=k_1\frac{m_d}{m_b}\ ;\ \ \ 
O_{23}^d=k_2\frac{m_s}{m_b} \ ,
\end{equation}
where we assume $k_i$ to be of order 1. We use analogous expressions for 
$O_{13}^u$ and $O_{13}^u$, keeping the same 
$k_i$ factors for simplicity. We
take the quark mass ratios in the ranges \cite{Hagiwara:fs}: 
\begin{equation}
\label{ran}
\begin{array}{ccc}
0.21\ \leq \ \left( \frac{m_d}{m_s}\right) ^{\frac 12}\leq 0.24 & \ ; & 
3.8\times 10^{-2}\ \leq \left( 
\frac{m_u}{m_c}\right) ^{\frac 12}\leq \ 7.2\times 10^{-2} \\  &  &  \\ 
2.1\times 10^{-2}\leq \ \frac{m_s}{m_b}^{}\leq \ 4.2\times 10^{-2}\  & ; & 
3.3 \times 10^{-3}\ \leq \ \frac{m_c}{m_t}^{}\ \leq \ 4.1 \times 10^{-3}
\end{array}
\end{equation}
Allowing $k_i$ to vary independently in the range 
\begin{equation}
\label{kik}0.8\ \leq k_i\ \leq \ 1.4\ \ \ i=1,2 \ ,
\end{equation}
and keeping the factorizable phases $\phi _1$, $\phi _3$ as free parameters,
we evaluate $\beta $ in Eq.~(\ref{beta}) from Eqs.~(\ref{lot}). We plot in
Fig.1 the allowed values for $\beta $ and the CP violation invariant 
$I=|{\rm Im}%
[V_{ij}V_{kl}V_{kj}^{*}V_{il}^{*}]|$. 
Note that all points included in Fig. 1 correspond to values of $%
|V_{us}|$, $|V_{cb}|$ and $|V_{ub}|/|V_{cb}|$ within the range
allowed by experiment, and with positive $\beta $. It is clear from Fig. 1
that, without non-factorizable phases, the central value for $\beta $ is
around $18^{\circ }$ and the following bound holds: 
\begin{equation}
\label{bet}\beta \leq 21^{\circ }\ ;\ \ \ \sin 2\beta \leq 0.67.
\end{equation}
This is to be compared with the combined BaBar and Belle results \cite
{updated}: 
\begin{equation}
\label{xxx}\sin (2\beta )=0.739\pm 0.048.
\end{equation}
It is clear that such low values for $\beta $ are already strongly
disfavoured (more than  $1 \sigma$ deviation) and will be ruled out by a more
precise measurement of $\sin (2\beta )$, provided the central value does not
decrease significantly. These low values of $\beta $ were to be expected
from our previous evaluation of the size of the leading contribution, given
in Eq.~(\ref{lea}). The values obtained for $I$ in Fig. 1 are also low with
respect to the experimental range given in \cite{Hagiwara:fs} ($I$ = $%
(3.0\pm 0.3)\times 10^{-5}$). It follows from our discussion that if one
works in a WB where the Yukawa matrices are Hermitian, with vanishing $(1,1)$
elements, and conforms to some naturalness requirements, one cannot obtain a
sufficiently large value of $\beta $, without allowing for non-factorizable
phases. This point is specially relevant since, as previously pointed out,
in a large class of anz\"atze considered in the literature, non-factorizable
phases cannot be introduced. The difficulty of obtaining a sufficiently
large value of $\sin (2\beta )$ in a specific ansatz of the above type was
recently pointed out in \cite{Kim:2004ki}. 
\begin{figure}
\centerline{\ \epsfysize=8.0truecm \epsffile{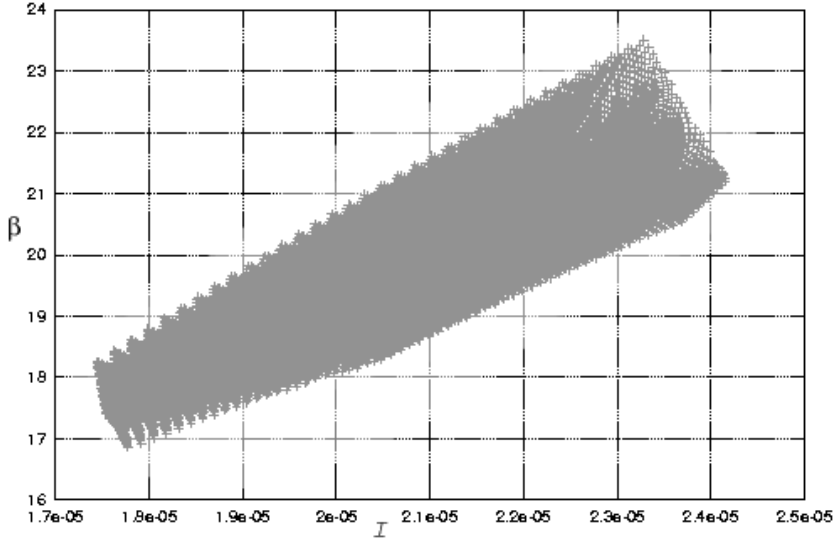}}
\caption{$\beta $ as a function of $I$, with non-factorizable phases
and central values for quark masses}
\label{bej3}
\end{figure}

We consider now the general case, allowing for the presence of
non-factorizable phases. The values of the elements of $V_{CKM}$ are now
given by expressions analogous to those in Eqs.~(\ref{lot}), with the
orthogonal real matrices $O^{u,d}$ substituted by the unitary matrices $%
U_o^{u,d}$, as defined in Eq.~(\ref{puk}). Note that $U_o^{u,d}$ now contain
non-factorizable phases $\sigma _u$, $\sigma _d$. For simplicity, we adopt
the ``standard parametrization'' \cite{Hagiwara:fs} for $U_o^u$, $U_o^d$. It
is then easy to understand why, with the presence of at least one
non-factorizable phase, one can obtain a sufficiently large size of $\sin
2\beta $. The reason is that, in the case of no non-factorizable phases, 
one has in leading order  
$\arg (V_{td})\simeq \arg (V_{tb})\simeq \phi _3$, and
therefore the leading contribution of $V_{td}$ to $\beta \equiv \arg
(-V_{cd}V_{tb}V_{cb}^{*}V_{td}^{*})$ just cancels that of $V_{tb}$. In the
presence of non-factorizable phases, this no longer happens since $%
(U_o^d)_{31}$ can have a significant phase. Figures 2 and 3 clearly
illustrate this point. In these figures, once again, we allow the $k_i$ to
vary within the range given by Eq.~(\ref{kik}), and keep $\phi _1$ as a free
parameter while fixing now $\phi _3=0$ for simplicity. Furthermore the
phases $\sigma _u$ and $\sigma _d$ are taken to be equal to each other and
allowed to vary freely. The difference between Fig. 2 and Fig. 3 is due to
the fact that, in plotting Fig. 2, we did our calculations with the quark mass
ratios fixed in their central values as given by Eq.~(\ref{exp}), whilst in
Fig. 3 we covered the whole allowed range as given by Eq.~(\ref{ran}).
Although we have set $\phi _3=0$, it should be pointed out that $\beta $ is
almost unaffected by variations in the value of $\phi _3$. Comparing Figs.
2 and 3, we conclude that in both cases sufficiently large values of $\beta $
can be obtained. The choice of $\sigma _u$ equal to $\sigma _d$ follows our
rationale of treating on an equal footing the up and down quark sectors as was
already done for the variables $k_i$. However, the value of $\beta $ is not
sensitive to this choice, since it is $\sigma _d$ that plays the crucial
r\^ole in generating, together with $\phi _1$, a sufficient large value for $%
\beta $. 
\begin{figure}
\centerline{\ \epsfysize=8.0truecm \epsffile{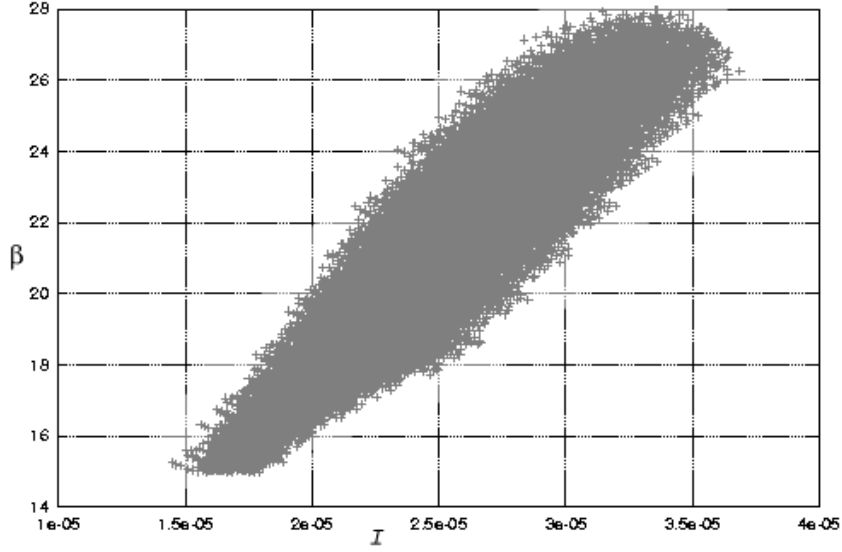}}
\caption{$\beta $ as a function of $I$, with non-factorizable phases
and quark masses varying within the allowed range}
\label{bej1}
\end{figure}

The reconstruction of the mass matrices from the experimental data is then
possible and one obtains, to leading order: {\small 
\begin{equation}
\label{rec}|M_u| \sim m_t\left( 
\begin{array}{ccc}
0 & \frac{\sqrt{m_um_c}}{m_t} & k_1
\frac{m_u}{m_t} \\ \frac{\sqrt{m_um_c}}{m_t} & \frac{m_c}{m_t} & k_2
\frac{m_c}{m_t} \\ k_1\frac{m_u}{m_t} & k_2\frac{m_c}{m_t} & 1
\end{array}
\right) ;\quad |M_d| \sim m_b\left( 
\begin{array}{ccc}
0 & \frac{\sqrt{m_dm_s}}{m_b} & k_1
\frac{m_d}{m_b} \\ \frac{\sqrt{m_dm_s}}{m_b} & \frac{m_s}{m_b} & k_2
\frac{m_s}{m_b} \\ k_1\frac{m_d}{m_b} & k_2\frac{m_s}{m_b} & 1
\end{array}
\right) \quad 
\end{equation}}
The structure of these mass matrices in terms of powers of $\varepsilon$
and $\overline \varepsilon $ is given by: 
\begin{equation}
\label{lll} |M_u| \sim m_t\left( 
\begin{array}{ccc}
0 & \varepsilon ^3 & \varepsilon ^4 \\ 
\varepsilon ^3 & \varepsilon ^2 & \varepsilon ^2 \\ 
\varepsilon ^4 & \varepsilon ^2 & 1
\end{array}
\right) \quad ;\quad |M_d| \sim m_b\left( 
\begin{array}{ccc}
0 &\overline \varepsilon   ^3 & \overline \varepsilon  ^4 \\ 
\overline \varepsilon  ^3 & \overline \varepsilon  ^2 & 
\overline \varepsilon  ^2 \\ 
\overline \varepsilon  ^4 & \overline \varepsilon  ^2 & 1
\end{array}
\right) \ ,
\end{equation}
with $\varepsilon = 0.06$ and $\overline \varepsilon = 0.2$.
This Froggatt--Nielsen pattern coincides with the ansatz in \cite
{Branco:1999tw} and also with that
in \cite{Roberts:2001zy}. The above analysis
illustrates the importance of a precise measurement of $\sin (2\beta )$ in
obtaining restrictions on the allowed patterns for Yukawa couplings. A
measurement of $\gamma $ will also have a significant impact in obtaining
further constraints on the Yukawa structures, specially since 
$3\times 3$ unitarity leads to the exact relation \cite{Botella:2002fr}: 
\begin{equation}
\label{bot}\frac{|V_{ub}|}{|V_{cb}|}=\frac{|V_{cd}|}{|V_{ud}|}\ \frac{\sin
(\beta )}{\sin (\beta +\gamma )} \ ,
\end{equation}
which will further restrict the allowed range for the ratio $%
|V_{ub}|/|V_{cb}|$.

\section{Weak basis with nine texture zeros}

\setcounter{equation}{0} 
In this section, we address the question of finding
the maximal number of zeros that can be obtained starting with arbitrary
complex mass matrices $M_u$, $M_d$ and making WB transformations, without
imposing the requirement that mass matrices be Hermitian. It is important to
recall that, in the SM, Yukawa couplings are not required to be Hermitian. In
fact, within the framework of the SM (or $SU(5)$) it is very difficult (if
not impossible) to have a symmetry that automatically constrains the Yukawa
matrices to be Hermitian. On the other hand, it is well known that Hermitian
(or symmetric) Yukawa matrices can be obtained in the framework of
left--right-symmetric theories or within $SO(10)$ GUTs. 

Here, we show that
there is a WB with nine texture zeros and this is the maximal number of
zeros that can be obtained. In this WB, the number of 
free parameters (nine real
numbers and one phase) equals the number of physical quantities (six quark
masses plus the four physical parameters contained in $V_{CKM}$) and thus it
is possible to obtain a full reconstruction of $M_u$, $M_d$ from
the experimental input. In order to prove our result we start 
with a basis where 
$M_u$ is diagonal real and $M_d$ is a general matrix. We then proceed
through the following steps:

i)We first perform a unitary transformation on the right of $M_d$, leading
to two zeros: 
\begin{equation}
M_d\longrightarrow M_d\cdot U=\left( 
\begin{array}{ccc}
0 & \times & \times \\ 
\times & \times & \times \\ 
0 & \times & \times 
\end{array}
\right) \ ;
\end{equation}
this is achieved by choosing the first column of $U$ orthogonal to the first
and third rows of $M_d$. Since this is a transformation on the right, it is
not felt by $M_u$.

ii) The second step is again a unitary transformation on the right of $M_d$,
leading to an additional zero in the $(22)$ entry: 
\begin{equation}
M_d\longrightarrow M_d\cdot U^{(23)}=\left( 
\begin{array}{ccc}
0 & \times & \times \\ 
\times & 0 & \times \\ 
0 & \times & \times 
\end{array}
\right) \ ,
\end{equation}
where $U^{(23)}$ denotes a three-by-three unitary matrix acting only in the $%
(23)$ space. Of course the first column of $M_d$ remains unchanged.

iii) The third step is a unitary transformation on the left which, of
course, will also have to be applied to $M_u$, in order to keep charged weak
currents diagonal real 
\begin{equation}
M_d\longrightarrow U^{(13)}\cdot M_d=\left( 
\begin{array}{ccc}
0 & \times & 0 \\ 
\times & 0 & \times \\ 
0 & \times & \times 
\end{array}
\right) . 
\end{equation}
The first row of the unitary matrix is chosen in such a way that a zero in
the $(13)$ entry of $M_d$ is generated. Applying the same unitary
transformation on the left of the diagonal matrix $M_u$, we obtain: 
\begin{equation}
M_u\longrightarrow \left( 
\begin{array}{ccc}
\times & 0 & \times \\ 
0 & \times & 0 \\ 
\times & 0 & \times 
\end{array}
\right) \ ,
\end{equation}
which can be put in the form 
\begin{equation}
M_u=\left( 
\begin{array}{ccc}
0 & \times & \times \\ 
\times & 0 & 0 \\ 
0 & \times & \times 
\end{array}
\right) 
\end{equation}
through a permutation of columns that corresponds to a unitary
transformation applied on the right.

iv) Finally an additional zero in $(13)$ of $M_u$ can be generated by a
unitary transformation on its right: 
\begin{equation}
M_u\longrightarrow M_u\cdot U^{(23)}=\left( 
\begin{array}{ccc}
0 & \times  & 0 \\ 
\times  & 0 & 0 \\ 
0 & \times  & \times 
\end{array}
\right) .
\end{equation}
The final texture after following these steps is of the form 
\begin{equation}
\label{nin}M_d=\left( 
\begin{array}{ccc}
0 & \times  & 0 \\ 
\times  & 0 & \times  \\ 
0 & \times  & \times 
\end{array}
\right) \ ;\ \ \ \ \ M_u=\left( 
\begin{array}{ccc}
0 & \times  & 0 \\ 
\times  & 0 & 0 \\ 
0 & \times  & \times 
\end{array}
\right) .
\end{equation}
It is clear that phase redefinitions allow for the elimination of all phases
in $M_d$, while the freedom left to perform phase redefinitions on the right
of $M_u$ leaves this matrix with a single phase in the second column. This
results in a total of nine real parameters and one phase, and thus equals the
number of physical measurable quantities. This fact implies the possibility
of fully reconstructing the quark mass matrices from experimental data.
In order to do so, a convenient strategy is to start with 
\begin{equation}
\label{stt}
\begin{array}{l}
M_u=
{\rm {diag}}\left( \frac{m_u}{m_t},\frac{m_c}{m_t},1\right)\ m_t \\  \\ 
M_d=V_{CKM}\cdot {\rm {diag}} \left(
\frac{m_d}{m_b},\frac{m_s}{m_b},1 \right) \ m_b \ ,
\end{array}
\end{equation}
where both $M_u$ and $M_d$ are already written in terms of physical
quantities and then
to perform the necessary transformations following
step-by-step the path outlined above. 

In the final result, we then further interchange the second
and third columns in $M_u$: 
\begin{equation}
\label{mmm}M_d=m_b\ \left( 
\begin{array}{ccc}
0 & m_{12}^d & 0 \\ 
m_{21}^d & 0 & m_{23}^d \\ 
0 & m_{32}^d & m_{33}^d
\end{array}
\right) \quad ;\quad M_u=m_t\ \left( 
\begin{array}{ccc}
0 & 0 & m_{13}^u \\ 
m_{21}^u & 0 & 0 \\ 
0 & m_{32}^u & m_{33}^u
\end{array}
\right) \ . \quad 
\end{equation}
Thus, we ensure that the diagonal entry $(33)$ is larger, or of the same
order, as the non-zero off-diagonal entries. In a leading order 
approximation,
we obtain: 
\begin{equation}
\label{ddd}
\begin{array}{l}
m_{12}^d=-a\ 
\frac{m_s}{m_b}\ |V_{us}| \\ m_{21}^d=b\ 
\frac{\sqrt{m_dm_s}}{m_b} \\ m_{23}^d=-a^{\prime }\ 
\frac{m_s}{m_b}) \\ m_{32}^d=1/a^{\prime } \\ 
m_{33}^d=1/a
\end{array}
\quad ;\quad 
\begin{array}{l}
a=
\sqrt{1+(m_s/m_b)/\ |V_{cb}|} \\ a^{\prime }=
\sqrt{1+\ |V_{cb}|/(m_s/m_b)} \\ b=(m_d/m_s)^{1/2}/\ |V_{us}|
\end{array}
\end{equation}
and 
\begin{equation}
\label{uuu}
\begin{array}{l}
m_{13}^u=p\ 
\frac{m_s}{m_b}\ \ |V_{us}|\ (1-q\ e^{-i\delta }) \\ m_{21}^u=
\frac{m_c}{m_t} \\ m_{32}^u=\ 
\frac{-m_u}{m_t}\ [1+q^2-2\ q\cos (\delta )]^{-1/2} \\ m_{33}^u=1
\end{array}
\quad ;\quad 
\begin{array}{l}
p=(m_s/m_b)/\ |V_{cb}| \\ 
q=\left( \frac{|V_{ub}|}{|V_{us}V_{cb}|}\right) \ \left( \frac{V_{cb}}{%
m_s/m_b}\right) ^2
\end{array}
\end{equation}
The structure of the matrices thus reconstructed in terms of powers of 
$\lambda =0.212 $ is of the form: 
\begin{equation}
\label{pow}M_d\sim \frac{m_b}{\sqrt{2}}\left( 
\begin{array}{ccc}
0 & \lambda ^3 & 0 \\ 
\lambda ^3 & 0 & \lambda ^2 \\ 
0 & 1 & 1
\end{array}
\right) \quad ;\quad M_u\sim m_t\left( 
\begin{array}{ccc}
0 & 0 & \lambda ^3 \\ 
\lambda ^3 & 0 & 0 \\ 
0 & \lambda ^7 & 1
\end{array}
\right) \quad .
\end{equation}
A few comments are in order at this stage. The matrix $M_uM_u^{\dagger }$ is
block-diagonal and is thus diagonalized by a two-by-two rotation with only $%
(13)$ mixing. As a result, in this WB, $(12)$ and $(23)$ mixing in $V_{CKM}$
arises only from $M_d$. Notice that we have chosen a phase convention such
that the single CP-violation phase was placed in $(M_u)_{13}$. Furthermore,
one can choose a WB, given by Eq.~(\ref{nin}), where the quark mass matrices
have the NNI form: apart from the $(3,3)$ element, the only
non-vanishing entries are $(1,2)$, $(2,1)$, $(2,3)$, $(3,2)$. It has been
shown \cite{Branco:1988iq} that starting with arbitrary matrices $M_u$, $M_d$%
, one can always make WB transformations such that $M_u$, $M_d$, are put in
the NNI form. Here, we conclude, that within the NNI weak basis, it is 
possible to obtain an extra zero entry.
It is clear from Eq.~(\ref{pow}) that the experimental data lead to strong
deviations of hermiticity in $M_d$. This result was to be expected, taking
into account that hermiticity, together with the NNI form, leads to the
Fritzsch ansatz, which was ruled out once it was found that the top quark is
very heavy, with $m_c/m_t\ll m_s/m_b$. Recall that the Fritzsch ansatz
predicts $|V_{cb}| =| \sqrt{\frac{m_s}{m_b}}-e^{i\alpha }\sqrt{%
\frac{m_c}{m_t}} |$ and thus the experimental value of $|V_{cb}|$
could only be reproduced with a strong cancellation of the contributions to $%
V_{cb}$ arising from $(m_s/m_b)^{1/2}$ and $(m_c/m_t)^{1/2}$. With a large
top quark mass, these cancellations became impossible. The interesting point
of the above derivation of $M_d$ from experiment 
is that it shows that deviations
from hermiticity have to be strong in the $(2,3)$ sector with $%
|m_{23}^d|/|m_{32}^d|\sim \lambda ^2$, while one has $|m_{12}^d|\sim
|m_{21}^d|$. The fact that the maximal number of zeros is odd prevents the
treatment of $M_u$ and $M_d$ on an equal footing.

\section{Conclusions}
\setcounter{equation}{0} 
Using an Hermitian WB for Yukawa couplings, we have parametrized
the up and down quark mass matrices through Eqs.~(\ref{uud}), 
 (\ref{puk}), separating the phases appearing in $M_u$, $M_d$
into factorizable and non-factorizable. We have then chosen to work,
without loss of generality, in a WB where the  $(1,1)$ elements
of $M_u$, $M_d$ both vanish. It was then pointed out, that in
the framework of this WB, the existence of at least one
non-factorizable phase is crucial to generate,
in a natural way, a value of $\sin (2\beta )$ sufficiently large
to be in agreement with experiment. This result is specially
relevant in view of the fact that many of the Yukawa textures 
proposed in the literature have the $(1,1)$ zero, together
with another off-diagonal texture zero in both $M_u$, $M_d$
thus implying the absence of non-factorizable phases in a
large class of ans\" atze. From our analysis we 
conclude that the present experimental data and in particular
the value of  $\sin (2\beta )$  does not favour the existence
of simultaneous off-diagonal texture zeros in $M_u$, $M_d$, in the
$(1,1)=0$ WB.
Allowing for the presence of non-factorizable phases one can, of course,
obtain a value of $\sin (2\beta )$ in agreement with experiment. The
reconstruction of the Yukawa flavour structure from input data is then
possible in the $(1,1) = 0$ basis, provided one adopts some naturalness
requirements. From the experimental data, 
we also derive a Froggatt--Nielsen
pattern for the quark mass matrices. Finally it is worth emphasizing the
important r\^ole that a more precise measurement of $\sin (2\beta )$ and a
measurement of $\gamma $ will have in narrowing down  the allowed
Yukawa textures.

\section*{Acknowledgements}

The authors thank the CERN Theory Division for warm hospitality. This work
was partially supported by CERN and by Funda\c c\~ao para a Ci\^encia e a
Tecnologia (FCT) (Portugal) through the Projects POCTI/FIS/36288/2000,
POCTI/FNU/43793/2002, POCTI/ FNU/49489/2002 (which have a FEDER component)
and CFIF - Plurianual (2/91).

\end{document}